\documentclass[%
superscriptaddress,
prl,
preprint,
 amsmath,amssymb,
 aps,
]{revtex4-1}

\usepackage{graphicx}
\usepackage{dcolumn}
\usepackage{bm}
\usepackage{color}
\begin{document}

\title{Magnetization control by angular momentum transfer from surface acoustic wave to ferromagnetic spin moments}

\author{R. Sasaki}
\thanks{Present address:Center for Emergent Matter Science (CEMS), RIKEN, Wako 351-0198, Japan}
\affiliation{Department of Basic Science, University of Tokyo, Meguro-ku, Tokyo 153-8902, Japan
}
\author{Y. Nii}
\affiliation{PRESTO, Japan Science and Technology Agency (JST), Kawaguchi 332-0012, Japan}
\affiliation{
Institute for Materials Research, Tohoku University, Sendai 980-8577, Japan
}
\author{Y. Onose}

\affiliation{
Institute for Materials Research, Tohoku University, Sendai 980-8577, Japan
}


\maketitle
\textbf{
The angular momentum interconversion between electron spin and other type of angular momenta is useful to develop new spintronic functionalities \cite{Otani2017}. 
The conversions from the angular momentum of photon \cite{Kimel2005,Stanciu2007} and mechanical rotation \cite{Barnet1915} to ferromagnetic spin moment have been well studied.
While the recent studies theoretically suggested circular vibration of atoms works as angular momentum of phonon \cite{Zhang2014,Garanin2015,Ruckriegel2020,Hamada2020}, the direct experimental demonstration of conversion to spin moments remains to be performed.
Here we demonstrate that the phonon angular momentum of surface acoustic wave can control the magnetization of a ferromagnetic Ni film by means of the phononic-to-electronic conversion of angular momentum in Ni/LiNbO$_3$ hybrid device.
This result clearly shows the phonon angular momentum is useful to further fictionalize spintronic devices.
}

Angular momentum is conserved when the system has rotational symmetry. While this law is, strictly speaking, broken in crystals, approximate conservation seems still valid in microscale range. For example, when spin-polarized electric current is injected to a microscale ferromagent, the spin angular momentum of conduction electron is transferred to the ferromagnetic localized moment (Fig. 1a). This mechanism is used to control the magnetic memory in the magnetoresistive random access memory \cite{Brataas2012}. One might wonder whether it is possible to control the magentization by the angular momentum transfer from phonons (Fig. 1b). 

The phonon angular momentum is activated by the breaking of time-reversal or spatial-inversion symmetry \cite{Zhang2014,Zhang2015,Hamada2018,Holanda2018,Zhu2018,Chen2019}. In time-reversal symmetry broken ferromagnets, the polarization of transverse acoustic wave (low-energy phonon mode) is observed to rotate while propagating along the magnetization \cite{Matthews1962}, which indicates the eigenstate with the circular polarization.
Similar circular polarized phonon is also observed in spatial inversion symmetry broken chiral materials, corresponding to the phonon version of natural activity \cite{Pine1970}. 
The phonon angular momentum is also emergent on a surface of substance. 
A Rayleigh type surface acoustic wave (SAW) has elliptically polarized displacement \cite{landau1986theory}, which indicates that the phonon angular momentum is finite (Fig. 1c). The angular momentum is parallel to the vector product of the SAW wave vector $k$ and surface normal vector, and it shows sign change when $k$ is reversed \cite{Long2018}. Here we use the SAW current to demonstrate the conversion from phonon angular momentum to magnetization.

Figure 2a shows the SAW device used in this work. This device is composed of piezoelectric LiNbO$_3$ substrate, two interdigital transducers (IDTs) and ferromagnetic Ni film \cite{Weiler2011,Dreher2012,Sasaki2017,Tateno2020}. For explanation, $xyz$-coordinate system is defined as shown in the right panel. 
To understand the coupling between the SAW and ferromagnetism, we first demonstrate how the magnetization direction of the Ni film affects the SAW propagation.
Figures 2b,c,e,f show the SAW transmission in magnetic fields nearly along $x$-axis. The magnetic field angle $\phi$ is slightly tilted ($\phi = 2^\circ$) from the $x$-axis to $z$ direction in Figs. 2b,c while the tilted direction is reversed in Figs. 2e,f. The magnetic field increases from -400 mT to 400 mT (decreases -400 mT to 400 mT) during the measurements of Figs. 2b,e (Figs. 2c,f). 
Before discussing the SAW transmission, let us explain the variation of magnetization in the magnetic fields. The insets illustrate the expected magnetization direction in the magnetic field sweep. Considering shape anisotropy of the Ni film, the easy and hard axis is the $z$- and $y$-axis, respectively. In this case, the magnetization variation sensitively depends on the tilted direction and the sign of the magnetic field variation. In decreasing the magnetic field at $\phi = 2^\circ$ (Fig. 2b), the tilting angle of magnetization $\theta$ is positive, and the magnitude increases. At zero magnetic field, the magnetization points along $+z$ direction.  When the magnetic field changes its sign, the magnetic state with negative $\theta$ is more energetically stable. Therefore, the sign of $\theta$ is abruptly reversed at some negative field, which is denoted as $\theta$ flop. In increasing field (Fig. 2c), the magnetization points along $-z$ direction at zero magnetic field and the $\theta$ flop shows up at a positive magnetic field. 
For the $\phi = -2^\circ$ measurements (Figs. 2e,f), the magnetization shows similar variation but the sign of $\theta$ and the magnetization direction at zero field are opposite.

Then, let us move on to the SAW transmission. We measured the transmission intensity from IDT1 to IDT2 $T_{+k}(H)$ and that from IDT2 to IDT1 $T_{-k}(H)$ (see supplementary information for precise definition of $T_{+k}(H)$ and $T_{-k}(H)$). For all the measurements, $T_{+k}(H)$, and $T_{-k}(H)$ shows broad dip around $\pm$ 90 mT. This is ascribed to the ferromagnetic resonance (FMR) excitation by the acoustic wave via magnetoelastic coupling \cite{Weiler2011,Dreher2012}. The discontinuous changes at -70 mT in Figs. 2b, e and those at +70 mT in Figs. 2c,f are caused by the $\theta$ flops mentioned above. 
Importantly, the intensity of acoustically excited FMR depends on the propagation direction of SAW. We plot the difference of transmittance $T^{\rm NR}(H) = T_{+k}(H) - T_{-k}(H)$ at $\phi = +2^\circ$ and $-2^\circ$ in Figs. 2d and g, respectively. $T^{\rm NR}(H)$ is independent of the magnetic field sweep direction except for the region around the $\theta$ flop fields, but it shows the opposite sign when either the sign of field or $\phi$ is reversed. 
This phenomenon is denoted as nonreciprocal SAW propagation induced by the simultaneous breaking of time-reversal and spatial inversion symmetries \cite{Sasaki2017,Tateno2020}. In this case, the ferromagnetism and surface break the time reversal and spatial inversion symmetries, respectively. 
The nonreciprocity originates microscopically from the different polarization of $+k$ and $-k$ mode. As mentioned above, the SAW has elliptical polarization, and the rotational direction is reversed by the reversal of $k$. On the other hand, FMR can be excited only by the effective field with right handed circular polarization. These induce the difference of in the intensity of acoustic FMR between $+k$ and $-k$ SAWs.

Now we discuss the inverse effect of the nonreciprocal propagation. Intuitively, the inverse effect would be the control of time reversal symmetry or magnetization by using the spatial inversion symmetry broken surface state and the unidirectional SAW flux.  
To demonstrate this, we consider the magnetization variation in the field decreasing process after applying strong magnetic field along $x$-axis (Fig. \ref{fig:M control}a). In this case, the magnetization points along the $x$-axis at first, and then it is tilted to either $+z$ or $-z$ direction due to the shape anisotropy. The SAW flux along $+x$ or $-x$ direction is expected to control whether the magnetization is tiled to $+z$ or $-z$ direction. 

To confirm this conjecture we have performed a numerical simulation. The magnetization should vary following the Landau Lifshitz Gilbert equation expressed as
\begin{align}
\frac{d\bm{m}}{dt} = -\gamma \bm{m} \times \bm{H}_{\rm eff}+ \frac{\alpha}{M_s} \bm{m} \times \frac{d\bm{m}}{dt},
\end{align}
where $\bm{m}=(m_x,m_y, m_z)$, $\gamma$, $\alpha$, and $M_s$ are uniform magnetization vector, gyromagnetic ratio, Gilbert damping, and saturation magnetization, respectively. 
$H_{\rm eff}=(-\frac{\partial  F}{ \partial m_x},-\frac{\partial  F}{ \partial m_y},-\frac{\partial  F}{ \partial m_z})$ is the effective magnetic field, in which free energy $F$ is composed of Zeeman energy $F_{\rm Z}=-\bm{m} \cdot \bm{H}$ ($\bm{H}$ is external field), magnetic anisotropy $F_{\rm a}$, and magentoelastic field $F_{\rm me}$. For simplicity, we assume uniaxial magnetic anisotropy $F_{\rm a}=-Km_z^2$ ($K$ is constant) and magnetoelastic coupling energy for poly crystal given by \cite{Dreher2012,chikazumi2009physics}
\begin{align}
F_{\rm me} &= b \sum_i m_i^2 e_{ii} + b \sum_{i\neq j} m_i m_j e_{ij}.
\end{align}
$b$ is magneto-elastic coupling constants, $m_i$ and $e_{ij}$ is the $i$th component of magnetization and strain tensor, respectively. For Rayleigh-type SAW propagating along $x$ direction, $e_{xx}$, $e_{xy}$, and $e_{yy}$ tensors are present. Introducing $m_{\pm} = m_x \pm im_y$ and $e_{x\pm} = e_{xx} \pm 2ie_{xy}$, $F_{\rm me}$ can be reduced to
\begin{align}
F_{\rm me} &= \displaystyle\frac{1}{2}\Bigl[
b m_x(m_+ e_{x-} + m_- e_{x+})
+ 2 b m_y^2 e_{yy}
\Bigr].
\end{align}
This formula clearly shows the angular momentum transfer from the phononic to magnetic systems (see supplementary information for details).
Figure \ref{fig:M control}b shows the calculated magnetic field variation of $m_z$ under SAW current. 
We assumed SAW-induced dynamical displacement as purely circular $u_x(t) = u_0 \cos(kx-\omega t)$ and $u_y(t) = -{\rm sgn}(k)  u_0 \sin(kx-\omega t)$, and neglected decay along $y$ direction for simplicity. The sign of the rotational motion depends on the sign of wave vector $k$. Then the relevant strains are expressed as $e_{xx}(t) =\frac{\partial u_x}{\partial x}= -ku_0 \sin(kx-\omega t)$ and $e_{xy}(t) =\frac{1}{2}(\frac{\partial u_x}{\partial y} + \frac{\partial u_y}{\partial x})= -(|k|u_0/2) \cos(kx-\omega t)$. The other parameters used for numerical calculation is displayed in Methods.
At $t=0$, a strong magnetic field is applied along $x$-axis, and we assumed $\bm{m}=(M_s,0,0)$. Then, the magnetic field is slowly decreased as $H_0\exp(-t / t_0)$ with $t_0= 1\ \mu$s. 
$m_z$ evolves below around the anisotropy field $2M_sK$. 
Importantly, the $m_z$ direction depends on whether SAW polarization is clockwise (CW) or counter-clockwise (CCW), which correspond to the SAWs propagating along $+x$ and $-x$ directions, respectively. 
This demonstrates deterministic control of magnetization by selecting SAW direction. 
The mechanism seems to be the damping torque dependent on the direction of polarization rotation. The circular polarization induce the rotational motion of magnetization in $xy$ plane. The damping torque due to the rotational motion $\tau_z=(\bm{m}\times d\bm{m}/dt)_z$ is parallel or antiparallel to the $z$-axis, and the sign depends on the direction of rotation.
Note that these results are independent on the phase of SAW (see supplementary information) and thus is different from precessional switching \cite{Thevenard2016}. 

Then, we present the corresponding experimental demonstration of the  magnetization control by the phonon angular momentum. We first applied the magnetic field as large as 400 mT along the $x$-axis and the SAW current along either $+x$ or $-x$ direction (From IDT1 to IDT2 or IDT2 to IDT1), of which the excitation power and frequency are 25 dBm and 2.906 GHz, respectively. To be precise, the magnetic field seems to be slightly tilted to the $+z$ direction but the angle between the magnetic field and the $x$-axis is less than $0.5^\circ$ (see Supplementary information). Then we decreased the magnetic field to zero at a rate of 0.01 T$/$s. Hereafter, we call the sequence of these operations \lq\lq poling procedure\rq\rq.
To detect the magnetization direction after the poling procedure, we used the magnetoresistance. Figure \ref{fig:M control}c shows the magnetoresistance $R(H)$ measured in magnetic fields $H$ parallel to the electric current in the Ni film. 
It shows a butterfly shape hysteresis loop, which corresponds to a typical magnetization curve with a finite coercive force (Fig. \ref{fig:M control}d). Therefore one can distinguish the magnetization state at zero magnetic field ($\bm{m}||+z$ or $\bm{m}||-z$) by measuring the magnetoresistance along $z$-axis. When the resistance decreases with increasing magnetic field from $H=0$ and shows a discontinuously increases at some magnetic field, the magnetization should have pointed to $-z$ direction at zero field. On the other hand, when it increases continuously without any discontinuity, the magnetization direction is opposite. If the magnetic field is decreased from $H = 0$, the field dependence for the magnetic states of $\bm{m}||+z$ and $\bm{m}||-z$ should be reversed. To probe the magnetization in this way, 
we rotated the device by 90$^\circ$ around $y$-axis after the poling procedure and measured the field dependence of the resistance along the $z$-axis with increasing or decreasing the field from 0 mT.

Figures \ref{fig:M control}e and f show the magnetic field dependence of $\Delta R(H) = R(H) - R(0)$ after the field poling with SAW current along $+x$ and $-x$ directions, respectively.
In the case of the SAW current along $+x$, the resistance increases with increasing the magnetic field from 0 mT almost continuously while it decreases at first with decreasing the field from 0 mT and shows a discontinuous increase around -10 mT. Conversely, after the poling with the SAW current along $-x$, the resistance decreases (increases) with increasing (decreasing) magnetic field from 0 mT. A discontinuous increase of resistance is observed only when the magnetic field is decreased. 
These results demonstrate that the SAW currents along $+x$ and $-x$ align the magnetization along $-z$ and +$z$, respectively. 
Thus, the control of magnetization by means of SAW current is certainly realized. The magnetization direction is opposite to the phonon angular momentum direction, which is consistent with the numerical simulation as shown in Fig. \ref{fig:M control}b.
For more detail on the input power and angle $\phi$ dependence, see Supplementary information.

To confirm the experimental demonstration of  magnetization control, we also probed the magnetization after the poling procedure by using nonreciprocal SAW transmissions $T^{\rm NR}(H)$. 
Figures \ref{fig:M control}g and f show $T^{\rm NR}(H)$ in increasing and decreasing the magnetic field almost along $x$-axis, respectively. 
The angle of magnetic field $\phi$ is less than $0.5^\circ$  but, precisely speaking, seems to slightly deviate from zero to positive side because the magnetic field dependence of $T^{\rm NR}(H)$ is similar to the case of $\phi=2^\circ$. The magnetic hysteresis becomes larger, and the discontinuous sign change is overlapped with the dip or peak. Therefore, one can probe the magnetization at zero magnetic field by the magnetic field dependence of $T^{\rm NR}(H)$. If $T^{\rm NR}(H)$ shows a simple dip as the magnetic field is increased from zero, the magnetization direction at 0 mT was along +$z$. On the other hand, if $T^{\rm NR}(H)$ shows a peak followed by discontinuous sign change, the magnetization pointed $-z$  at zero field.
Figures \ref{fig:M control}i and f show $T^{\rm NR}(H)$ after the $H$ poling with SAW currents along $+x$ and $-x$ directions at the same angle, respectively. As shown in Fig. \ref{fig:M control}i, $T^{\rm NR}(H)$ shows peak and a sign change around 80 mT after poling with SAW current along $+x$. Therefore, the magnetization pointed $-z$ direction at $H=0$.
On the other hand, $T^{\rm NR}(H)$ after poling with SAW current along $-x$ shows a simple dip, indicating that the magnetization pointed along $+x$ at $H=0$. These results are consistent with the magnetoresistance measurements.
The same measurement at small negative angle also performed, and it exhibited the same result (see Supplementary information).

Thus, we have demonstrated the magnetization control by the angular momentum transfer from SAW to ferromagentic spin moments.
Nevertheless, it should be noted that the volume fraction of controlled magnetization seems to be less than 100\% in the experiments. The magnitude of resistance discontinuities in Figs. 3e and 3f are nearly 40 \% of that in Fig 3c. The magnetic field variations shown in Figs. 3i and j are weaker than those in Figs. 3g and h. The  volume fraction of the controlled ferromagnetic domain seems to be several tens percent.
A number of related experimental and theoretical studies were already reported. While the interconversion between the mechanical rotation and spin moment are known as Einstein-de Haas effect and Barnet effect \cite{einstein1915experimenteller,Barnet1915}, the present result demonstrates the conversion from the  angular momentum of the microscopic phonon excitation. More recently, Kobayashi \textit{et al.} reported the generation of alternating spin current from SAW \cite{Kobayashi2017}. This phenomenon is qualitatively different from the present result, which originates from the time-independent angular momentum in SAW. 
The concept of phonon angular momentum is theoretically developed in the spintronics field, recently \cite{Zhang2014,Garanin2015,Ruckriegel2020,Hamada2020,Zhang2014,Zhang2015,Hamada2018,Holanda2018,Zhu2018,Chen2019}. The present result experimentally demonstrates an important functionality as angular momentum, the conversion to the ferromagnetic spin moments, which prove the validity of phonon angular momentum. The functionality as angular momentum might be useful to further develop spintronic devices. In addition, the concept of angular momentum might pave the avenue between the acoustic device and spintronics because the SAW device is indispensable in the contemporary telecommunication technology.

\section{Methods}
The SAW device in this work was fabricated by the electron beam lithography technique. The substrates of the device is Y-cut LiNbO$_3$ and the SAW propagation direction is along Z-axis of the substrate. The IDTs and electrodes are made of Ti (5 nm) and Au (20 nm). One IDT has 200 pairs of 100 $\mu$m fingers, and the distance between two IDTs is 500 $\mu$m. 
The finger width and spacing of IDTs are both 300 nm. The corresponding wavelength and frequency are 1.2 $\mu$m and 2.9 GHz, respectively. 
The Ni film was sputtered between two IDTs on the LiNbO$_3$ substrate and connected to six electrodes for the resistance measurement. The thickness, width, and length are 30 nm, 10 $\mu$m, 175 $\mu$m, respectively.
After the Ni film sputtered, the device was kept at 200 $^\circ$C for 30 minutes to eliminate the strains of the Ni film that arise from the sputtering process. 

All the measurements in this work were done at 100 K.
The SAW transmission was measured by a vector network analyzer. The microwave power was 10 dBm in Figs. 2b-g and 3g,h, while that was -10 dBm in Figs. 3i and j.
The magnetoresistance was measured by a lock-in amplifier with the frequency of 11.15 Hz.

LLG equation was numerically solved by using Mathematica. We put realistic values into the coefficients used in the calculation. The  Saturation magnetization and Gilbert damping coefficient were $M_s = $370 kA/m$^3$ \cite{chikazumi2009physics} and $\alpha = 0.064$ \cite{Platow1998}, respectively. Magnetic anisotropy constant was assumed to be $K M_s^2 = 1.63 \times 10^4 $ N/m$^2$ for reproducing flipping magnetic field of $2 K M_s\simeq 0.088$ T. Strain amplitude and frequency of SAW were  $e_0 = |k| u_0 = = 2 \times 10^{-6}$ and  $f = \omega/2\pi = 2.9$ GHz, respectively. The magnetoelastic coupling constants $b$ were given by the average of $b_1$ and $b_2$, where $b_1 M_s^2 = 6.20 \times 10^6$ N/m$^2$ and $b_2 M_s^2 = 4.30 \times 10^6$ N/m$^2$, respectively \cite{chikazumi2009physics}.

\begin{acknowledgments}
\section{Acknowledgments}
The electron beam lithography was carried out by the collaborative research in the Cooperative Research and Development Center for Advanced Materials, Institute of Materials Research, Tohoku University with the help of K. Takanashi and T. Seki. This work was supported in part by JSPS KAKENHI grant numbers JP16H04008, JP17H05176, JP18K13494, 20K03828, PRESTO grant number JPMJPR19L6, the Mitsubishi foundation, and the Noguchi institute. 
R.S. is supported by the Grant-in-Aid for JSPS Research Fellow (No. 18J12130).
\end{acknowledgments}

\section{Author contributions}
R.S. and Y.O. conceived and designed the project and wrote the paper with the help of Y.N.
R.S. fabricated the device and performed all the measurement.
Y.N. performed theoretical calculation.
Y.O. supervised this project.

\section{Competing interests}
The authors declare no competing interests.

\bibliographystyle{naturemag}
\bibliography{bibliography}

\begin{figure}[t]
  \centering
  \includegraphics[width = 0.8\linewidth]{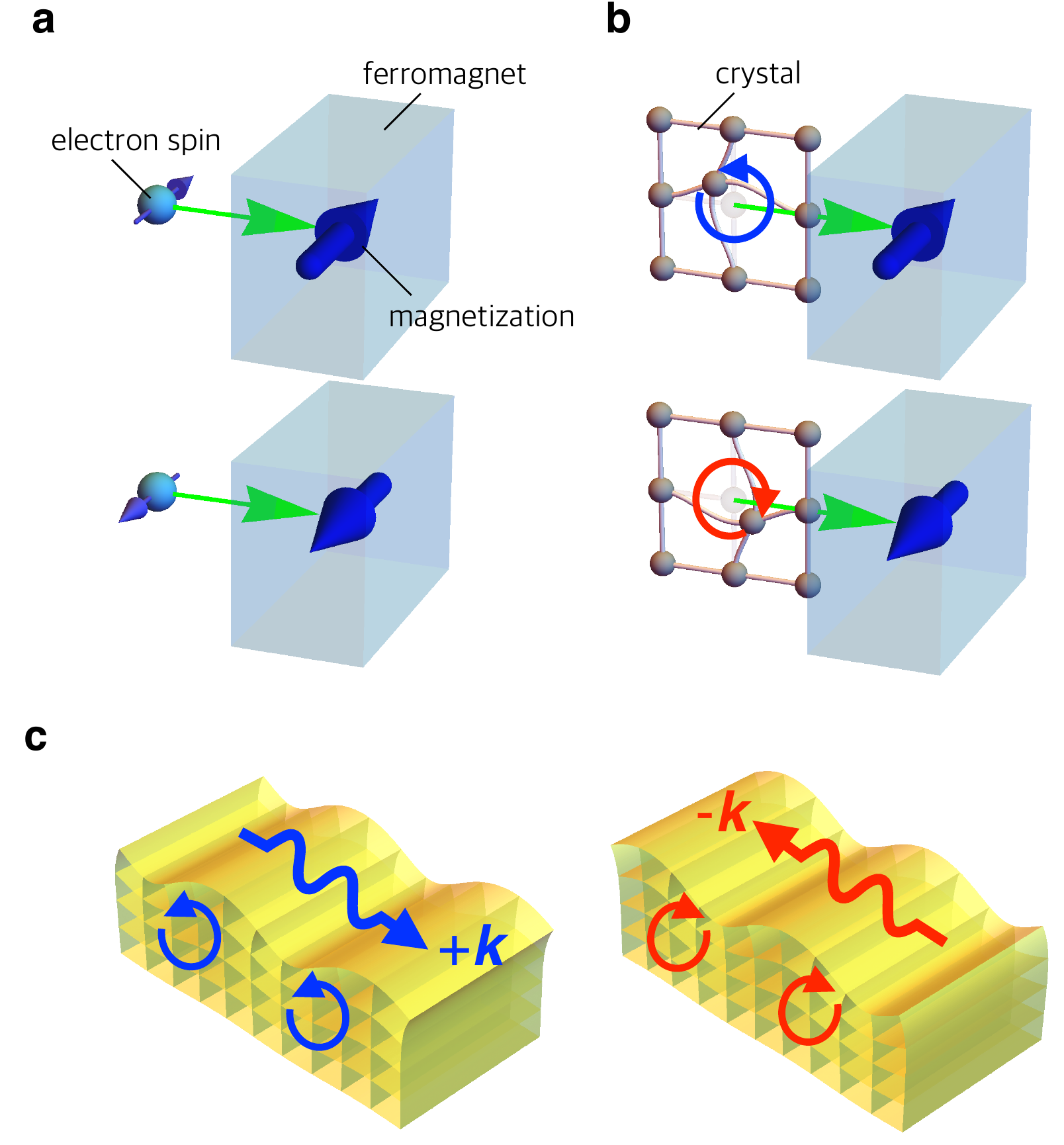}
  \caption{\textbf{Concept of magnetization control by phonon anglar momentum.} 
  \textbf{a}, Schematic illustrations of magnetization control by electron spin injection.
  \textbf{b}, Schematic illustrations of magnetization control by the injection of phonon angular momentum. 
  \textbf{c}, Schematic illustrations of phonon angular momenta in surface acoustic waves (SAWs). The direction of phonon angular momentum of SAW depends on the sign of wave vector $k$.
  }
  \label{fig:concept}
\end{figure}

\begin{figure}
  \centering
  \includegraphics[width = 0.5\linewidth]{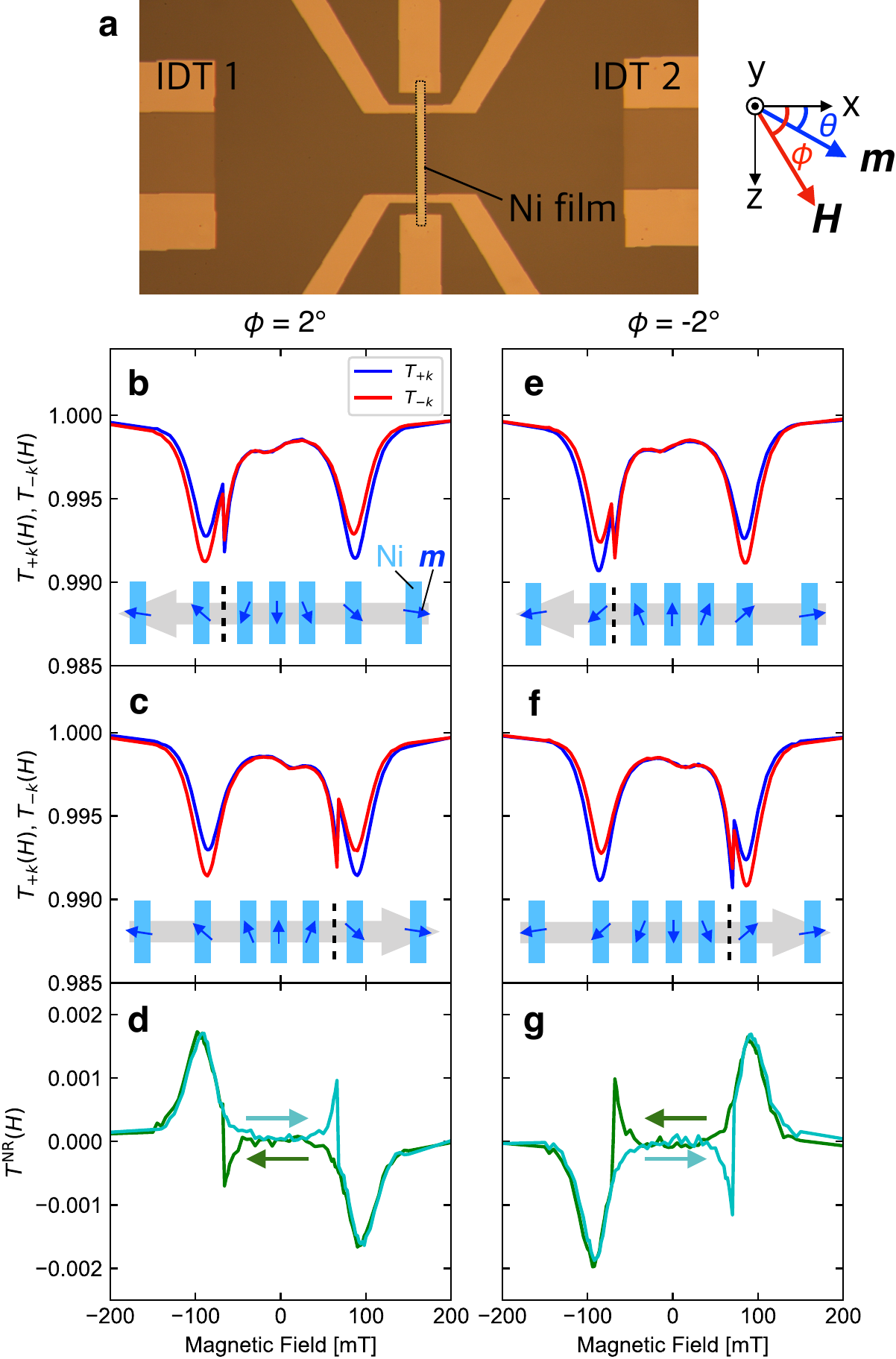}
  \caption{\textbf{Nonreciprocal SAW propagation.}
  \textbf{a}, An optical micrograph of the SAW device used in this work. Two interdigital transducers (IDT 1 and IDT 2) and Ni film with six electrodes are fabricated on LiNbO$_3$ substrate. The dotted rectangle emphasizes the Ni film. Right sketch illustrates $xyz$-coordinate system, the angle $\phi$ between the $x$-axis and magnetic field $\bm{H}$, and the angle $\theta$ between the $x$-axis and magnetization $\bm{m}$, which are used for the explanation.
  \textbf{b,c}, The magnetic field dependence of the SAW transmission along $+x$ and $-x$ directions ($T_{+k}(H)$, $T_{-k}(H)$) in decreasing (\textbf{b}) and increasing (\textbf{c}) field at $\phi = 2^\circ$. The insets illustrates the inferred magnetization directions in Ni film. Gray arrow and black dashed line represent the sweep direction of the magnetic field and the magnetization $\theta$ flop field, respectively.
  \textbf{d}, Nonreciprocity of the SAW transmission $T^{\rm NR}(H)=T_{+k}(H)-T_{-k}(H)$ in decreasing and increasing field at $\phi = 2^\circ$.
  \textbf{e,f}, The magnetic field dependence of the SAW transmission in decreasing (\textbf{b}) and increasing (\textbf{c}) the field at $\phi = -2^\circ$.
  \textbf{g}, $T^{\rm NR}(H)$  in decreasing and increasing field at $\phi = -2^\circ$.
  }
  \label{fig:nonreciprocity}
\end{figure}

\begin{figure}
  \centering
  \includegraphics[width = 0.5\linewidth]{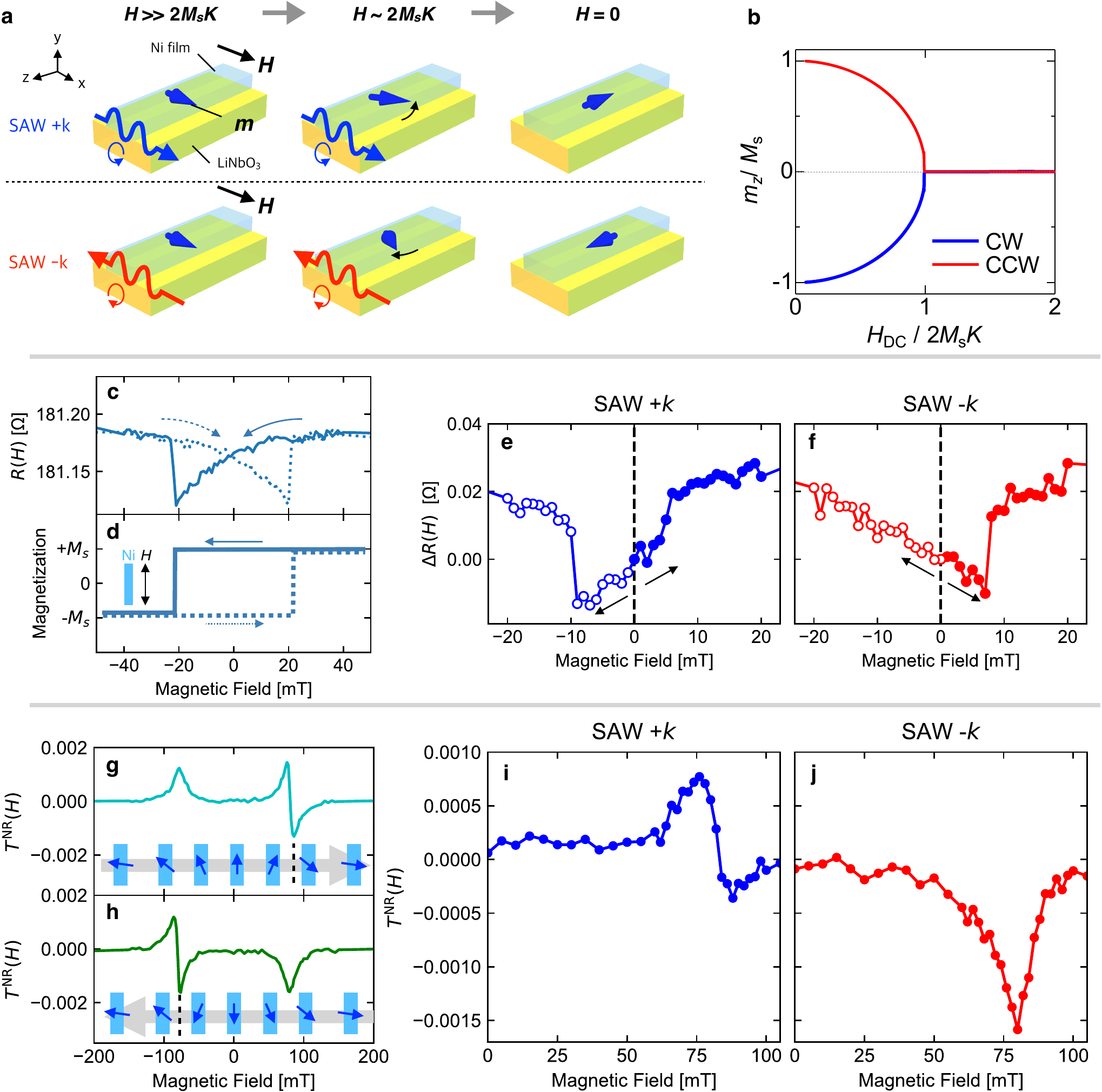}
  \caption{\textbf{Demonstration of magnetization control by phonon angular momentum.}
  \textbf{a}, Schematic illustrations of the magnetization $m$ control by phonon angular momentum of SAW. Here, $2M_sK$ is anisotropy field. 
  The application of SAW current with the wave vector parallel ($+k$) or anti-parallel ($-k$) to the $x$-axis (CW polarization or CCW polarization) during the decrease of magnetic field controls the magnetization direction at $H$ = 0.
  \textbf{b}, The numerically calculated  magnetization evolution under the application of SAW currents along $+x$ and $-x$. 
  \textbf{c}, The magnetoresistance of Ni thin film as a function of magnetic field along the $z$-axis ($\phi$ = $-90^\circ$).
  \textbf{d}, The illustration of magnetization curves inferred by the resistance measurement shown in \textbf{c}.
  \textbf{e,f}, The magnetic field dependence of the resistance difference $\Delta R(H) = R(H) - R(0)$ at $\phi = -90^\circ$ after the poling procedure with SAW currents along $+x$ (\textbf{e}) and $-x$ (\textbf{f}) directions. The dashed lines and the arrows indicate the initial field (0 mT) and field sweep directions, respectively.
  \textbf{g,h}, The magnetic field dependence of the SAW transmission nonreciprocity $T^{\rm NR}(H) = T_{+k}(H) - T_{-k}(H)$ in decreasing (\textbf{g}) and increasing (\textbf{h}) field. The magnetic field direction is very close to the $x$-axis. The angle $\phi$ is positive and the magnitude is less than 0.5$^\circ$ (see text). The insets illustrate the magnetization direction in the film inferred by the nonreciprocity. Gray arrow and black dashed line represent the sweep direction of the magnetic field and the magnetization $\theta$ flop field, respectively. 
  \textbf{i,j}, The magnetic field dependence of $T^{\rm NR}(H)$ measured with increasing magnetic field from 0 mT after the magnetic poling with SAW current along $+x$ (\textbf{i})  and $-x$ (\textbf{j}) directions.
  }
  \label{fig:M control}
\end{figure}

\end{document}


\title{Supplementary information for \lq\lq Magnetization control by angular momentum transfer from surface acoustic wave to ferromagnetic spin moments\rq\rq}

\author{R. Sasaki}
\thanks{Present address:Center for Emergent Matter Science (CEMS), RIKEN, Wako 351-0198, Japan}
\affiliation{Department of Basic Science, University of Tokyo, Meguro-ku, Tokyo 153-8902, Japan
}
\author{Y. Nii}%
\affiliation{PRESTO, Japan Science and Technology Agency (JST), Kawaguchi 332-0012, Japan}
\affiliation{
Institute for Materials Research, Tohoku University, Sendai 980-8577, Japan
}
\author{Y. Onose}

\affiliation{
Institute for Materials Research, Tohoku University, Sendai 980-8577, Japan
}

\maketitle

\setcounter{figure}{0} \renewcommand{\thefigure}{S\arabic{figure}}

\section{Theory for angular momentum transfer from phonon to spin moment}
In electronic systems, the spin-orbit interaction is written by $\mathcal{H}_{SOC} = \lambda (L_x S_x + L_y S_y + L_z S_z) = \lambda [(L_+ S_- + L_- S_+)/2 + L_z S_z)]$, where $L_{\pm}S_{\mp}$ terms represent the exchange of angular momentum between orbital and spin sectors. Here we show that Eq. (3) in the main text describes the exchange of angular momentum between a phonon and spin moment. In order to show this, we demonstrate that $e_{x+}$ and $e_{x-}$ operators raise or lower the phonon angular momentum, respectively.

For simplicity, we consider a system with a single atom in the unit cell and introduce the second quantization of the displacement vector in the $l$-th unit cell by
\begin{align*}
\bm{u}_l = 
\sum_k \sqrt{\displaystyle\frac{\hbar}{2\omega_{k} N}}
\Bigl(\bm{\epsilon}_ka_k e^{i(\bm{k} \cdot \bm{r}_l-\omega t)} + h.c.\Bigr),
\end{align*}
where $\bm{\epsilon}_k = (\epsilon_k^x, \epsilon_k^y, \epsilon_k^z)$ is the polarization vector, $a_k$ and $a_k^\dagger$ are phonon annihilation and creation operators for a mode $k = (\bm{k}, \sigma)$ with a wave vector $\bm{k} = (k_x, k_y, k_z)$ and a branch $\sigma$, respectively.
Then, the phonon angular momentum $\bm{J}_{ph} = \sum_{l}\bm{
u}_l \times\bm{\dot{u}}_l$ of $z$ component is given by \cite{Zhang2014}
\begin{align}
J_{ph}^z = \sum_k l_k^z
\Bigl[
n_k+ \displaystyle\frac{1}{2}
\Bigr].
\end{align}
Here $l_k^z = \hbar \bm{\epsilon}_k^\dagger M\bm{\epsilon}_k$ and $n_k = a_k^\dagger a_k$ represent phonon angular momentum and the number operator of phonons of an eigenmode specified by $k$, respectively, and
$M = \begin{pmatrix}
0 & -i & 0\\
i & 0 & 0\\
0 & 0 &0
\end{pmatrix}$.

Here we consider Rayleigh-type SAW propagating on $y$ surface along $\pm x$ direction. The phonon number states of these SAW with $\bm{k} = (\pm k, 0, 0)$ can be expressed as $\ket{n_{+k}}$ and $\ket{n_{-k}}$, respectively. Here we assume these SAW states has perfectly circular polarization, where $\bm{\epsilon}_{\pm k} = (1, \pm i, 0)/\sqrt{2}$. Then, the phonon angular momentum of these states can be given by
\begin{align}
&\braket{n_{+k}|J_{ph}^z|n_{+k}} = \hbar (n_{+k} +1/2), \\
&\braket{n_{-k}|J_{ph}^z|n_{-k}} = -\hbar(n_{-k} + 1/2).
\end{align}

Next, let us consider the quantum operators $e_{x\pm}$. Since $e_{x\pm} = e_{xx} \pm i (2e_{xy}) = \partial_x u_x \pm i(\partial_x u_y + \partial_y u_x)$, these are written by using $\bm{u}_l$ as
\begin{align*}
e_{x\pm} &= \sum_k
\sqrt{\displaystyle\frac{\hbar}{2\omega_{k} N}}\Bigl[
\displaystyle\frac{ik_x}{\sqrt{2}} \Bigl(
\epsilon^{\pm}_ka_k e^{i(\bm{k} \cdot \bm{r}_l-\omega t)}
-\epsilon^{\mp*}_k a_k^\dagger e^{-i(\bm{k} \cdot \bm{r}_l-\omega t)}
\Bigr)\\
&\mp k_y\Bigl(
\epsilon_k^xa_k e^{i(\bm{k} \cdot \bm{r}_l-\omega t)}
-\epsilon_k^{x*} a_k^\dagger e^{-i(\bm{k} \cdot \bm{r}_l-\omega t)}
\Bigr)
\Bigr],
\end{align*}
where $\epsilon^{\pm}_k = (\epsilon_k^x \pm i\epsilon_k^y)/\sqrt{2}$ .

Thus, we get
\begin{align*}
e_{x+}\ket{n_{+k}} &= -\sqrt{\displaystyle\frac{\hbar}{2\omega_{0} N}}
\displaystyle\frac{ik}{\sqrt{2}}
e^{-i(k x-\omega t)}\sqrt{n_{+k}+1}\ket{n_{+k}+1}, \\
e_{x-}\ket{n_{+k}} &= \sqrt{\displaystyle\frac{\hbar}{2\omega_{0} N}}
\displaystyle\frac{ik}{\sqrt{2}}
e^{i(k x-\omega t)}\sqrt{n_{+k}}\ket{n_{+k}-1},\\
e_{x+}\ket{n_{-k}} &= -\sqrt{\displaystyle\frac{\hbar}{2\omega_{0} N}}
\displaystyle\frac{ik}{\sqrt{2}}
e^{-i(k x+\omega t)}\sqrt{n_{-k}}\ket{n_{-k}-1},\\
e_{x-}\ket{n_{-k}} &= \sqrt{\displaystyle\frac{\hbar}{2\omega_{0} N}}
\displaystyle\frac{ik}{\sqrt{2}}
e^{i(k x+\omega t)}\sqrt{n_{-k}+1}\ket{n_{-k}+1},
\end{align*}
where $\omega_{0}$ is a frequency of these SAW. This clearly represent the $e_{x\pm}$ operators raise or lower the total phonon angular momentum of $J_{ph}^z$ by $+\hbar$.

\clearpage
\section{Analysis of SAW transmission spectra}
\begin{figure}[h]
  \centering
  \includegraphics[width = \linewidth]{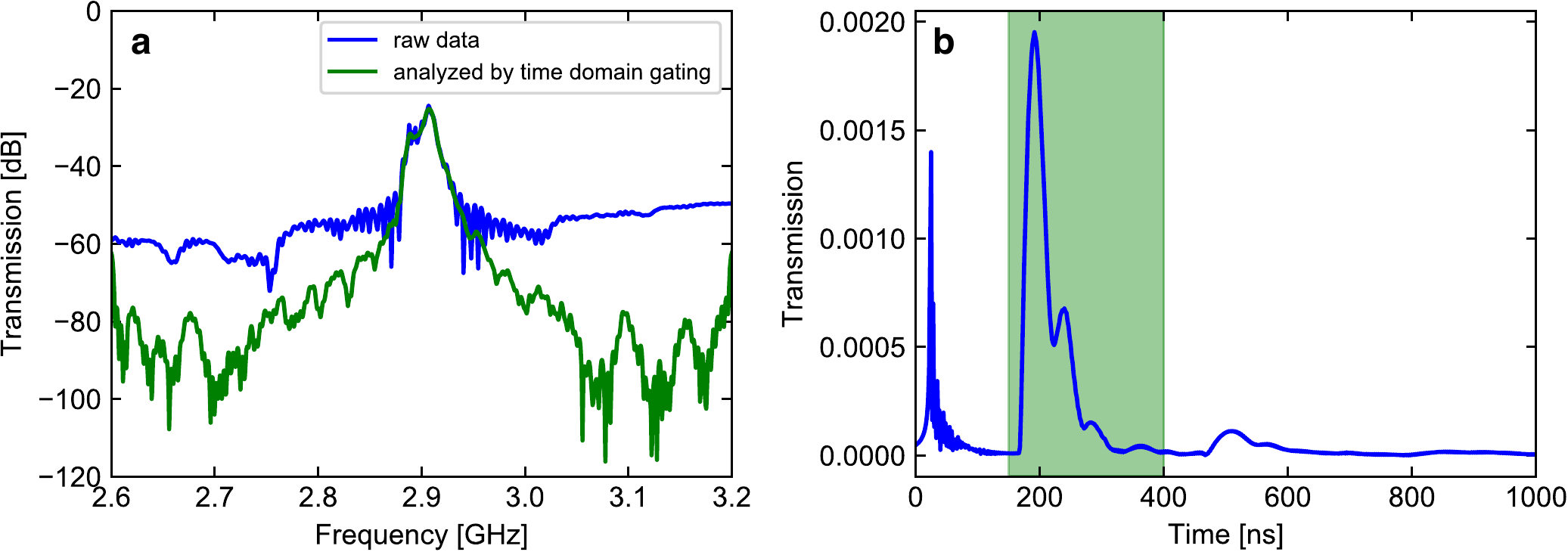}
  \caption{\textbf{Raw and analyzed SAW transmission spectra.}
  \textbf{a}, A raw transmission spectrum and the analyzed spectrum by using time domain gating.
  \textbf{b}, The transmission of the SAW device in time domain obtained by inverse Fourier transform.
  }
  \label{fig:SAW spectrum}
\end{figure}

Figure \ref{fig:SAW spectrum}a exemplified the absolute values of raw and analyzed transmission spectra from IDT 1 to IDT 2 ($S_{21}$) measured by a vector network analyzer. Not only the signal carried by SAW propagation but also the direct microwave crosstalk contribute to the raw spectra.  To remove the direct crosstalk contribution, we performed the time domain gating as follows.
By inverse Fourier transform of the transmission in frequency domain, we get the transmission as a function of time. Figure \ref{fig:SAW spectrum}b shows the absolute value of the transmission in time domain. The narrow peak at time near 0 ns is attributed to the direct microwave crosstalk between two IDTs. The transmission peaks around 170 ns are the main signal owing to SAW propagation. The small peak at 500 ns is caused by the multiple reflections from IDTs.
By Fourier transforming the time domain transmission only within the time region of the main SAW signal (150 - 400 ns), we obtained the SAW transmission spectrum (green line in Fig. \ref{fig:SAW spectrum}a), in which the direct crosstalk and the contribution of multiple reflection are removed.

The transmissions $T_{+k}(H)$ presented in the main text is defined as $T_{+k}(H)=\bar{S}_{21} (H)/\bar{S}_{21}({\rm400\ mT})$, where $\bar{S}_{21}$ is the average of time gated $S_{21}$ spectrum between 2.8 GHz and 3.0 GHz in the unit of V. We similarly obtained the $T_{-k}(H)=\bar{S}_{12} (H)/\bar{S}_{12}({\rm400\ mT})$, where $\bar{S}_{12}$ is the average of time gated $S_{12}$ spectrum.

\clearpage
\section{SAW transmission and nonreciprocity in the magnetic fields along various in-plane directions}
\begin{figure}[h]
  \centering
  \includegraphics[width = 0.75\linewidth]{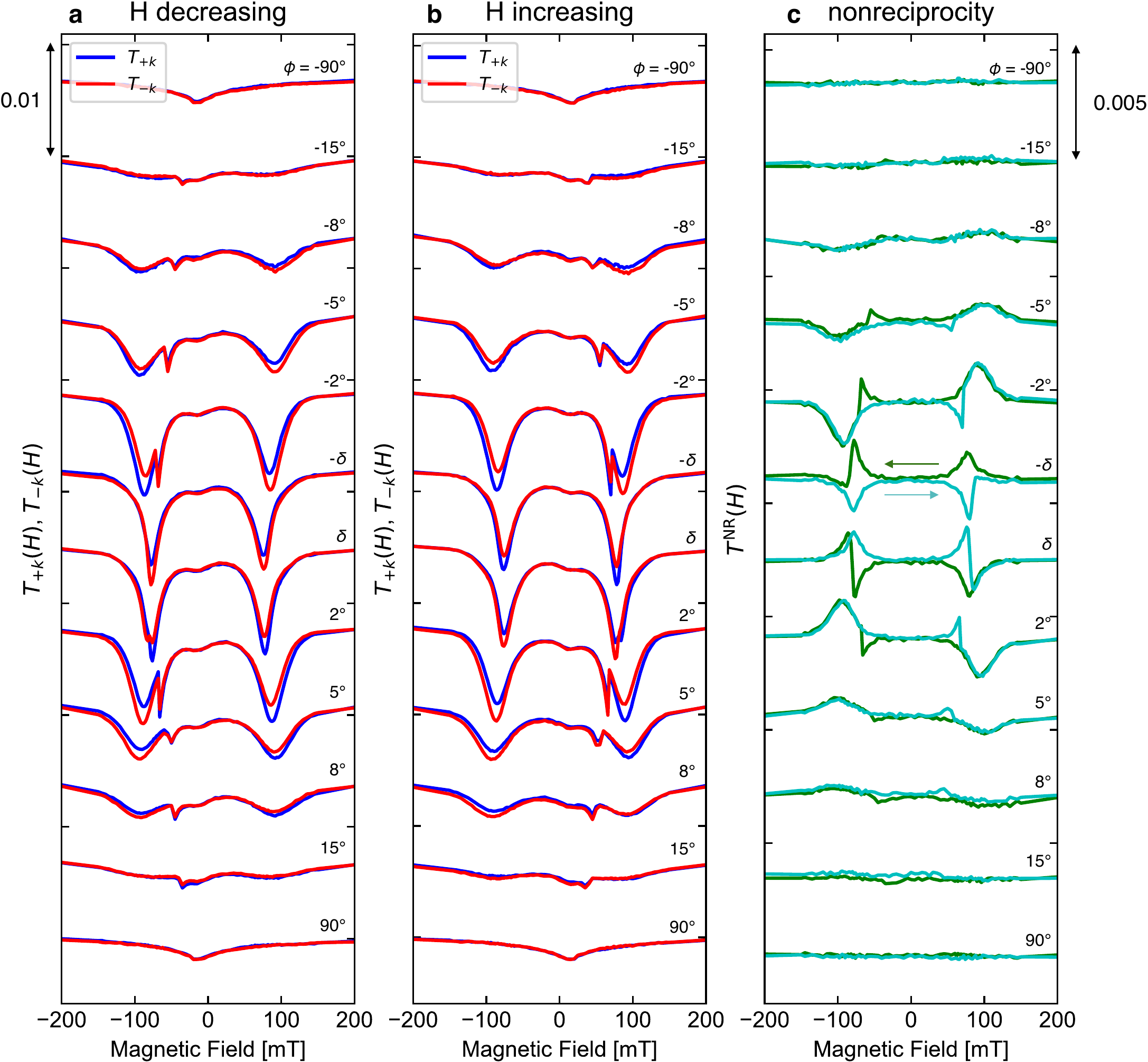}
  \caption{\textbf{SAW transmission and nonreciprocity in magnetic fields parallel to various in-plane directions}
  \textbf{a,b}, SAW transmission along $+x$ and $-x$ directions  $T_{+k}(H)$, $T_{-k}(H)$ in decreasing (\textbf{a}) and increasing (\textbf{b}) magnetic fields applied at various angles $\phi$.
  \textbf{c}, nonreciprocity  $T^{\rm NR}(H) =T_{+k}(H)-T_{-k}(H)$ in decreasing (green) and increasing (cyan) the magnetic field at various angle $\phi$.
  The  $+\delta$, $-\delta$ are positive and negative angles very close to 0 deg. The deviation is less than $0.5^\circ$. Initially, we adjust the rotator as close to zero as possible. Judging from the observed SAW transmission and magnetoresistance, the angle is slightly tilted to the positive side. $+\delta$ denotes this angle. After that, we rotated to the negative direction by $0.5^\circ$. Then, the angle seems to be slightly negative. We define this angel as $-\delta$. It is difficult to get $\phi$ closer to zero  reproducibly. Hereafter , we use these notations for the angles very close to zero.}
  \label{fig:T and NR}
\end{figure}

\begin{figure}
  \centering
  \includegraphics[width = 0.8\linewidth]{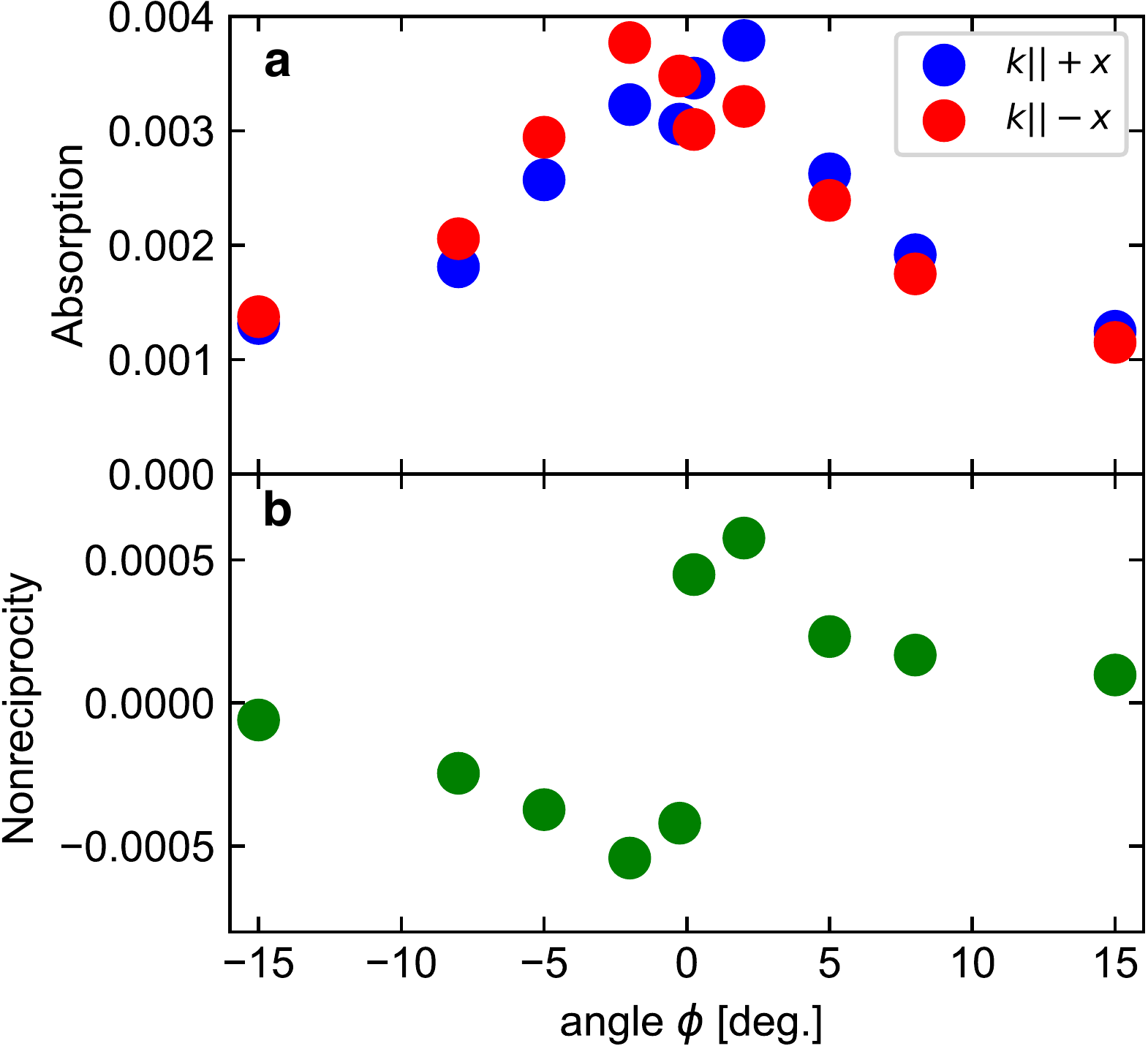}
  \caption{\textbf{The angle $\phi$ dependence of the SAW absorption and nonreciprocity.}
  \textbf{a}, $\phi$ dependence of the absorption of the SAW with the wave vectors $k$ parallel ($k ||+x$) and antiparallel ($k||-x$) to the $x$-axis in decreasing the magnetic field. The SAW absorption is calculated by averaging $-T_{+k}(H)$ and $-T_{-k}(H)$ in the region of $H>0$ for decreasing $H$ at each $\phi$.
  \textbf{b}, $\phi$ dependence of the nonreciprocity estimated by the difference of absorptions for $k||x$ and $k||-x$.}
  \label{fig:phi dep of SAW}
\end{figure}

\clearpage
\section{Phase dependence}
\begin{figure}[ht]
  \centering
  \includegraphics[width = \linewidth]{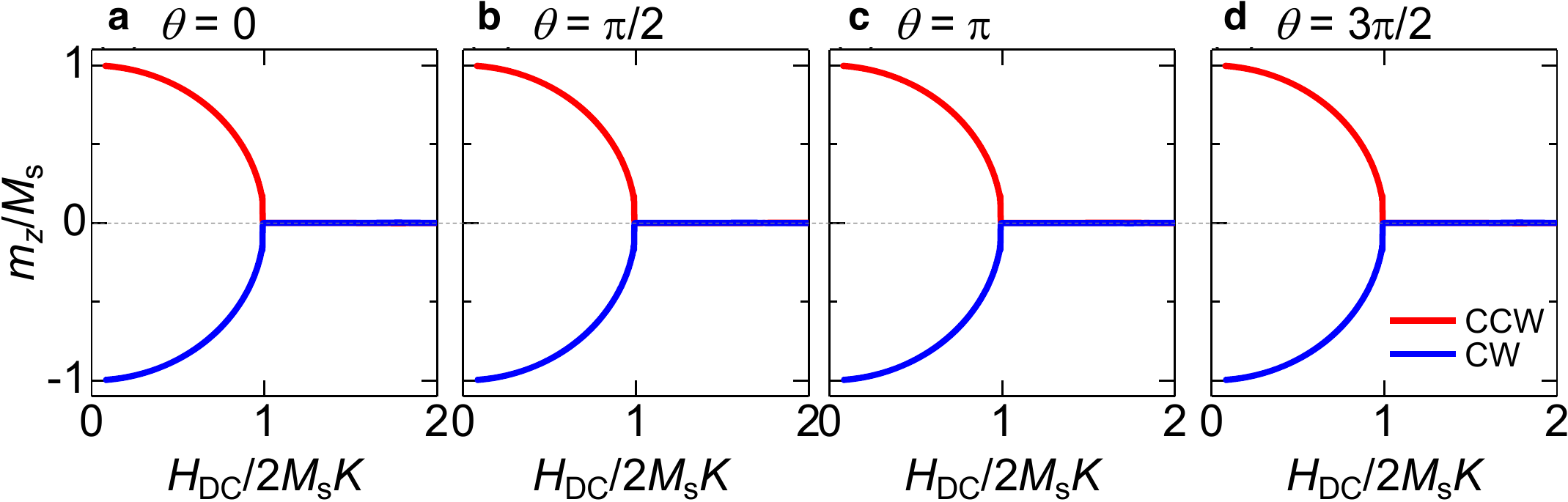}
  \caption{\textbf{SAW phase dependence of magnetization evolution under SAW poling.} 
  \textbf{a,b,c,d} The numerically calculated magnetic field dependence of magnetization under the application of SAW currents along $+x$ and $-x$ with various SAW initial phases $\theta$ = 0 (a), $\pi/2$  (b), $\pi$ (c), $3\pi/2$ (d).}
  \label{fig:Phase difference}
\end{figure}
The magnetization control in this work is totally different from the so-called precessional magnetization switching\cite{Thevenard2016}, in which the final magnetization direction depends on the phase of precession. To demonstrate the difference, we show in Fig. \ref{fig:Phase difference} the calculated magnetization evolution under the acoustic wave with clock wise (CW) and counter clockwise (CCW) polarization with various SAW initial phases $\theta$ at $t=0$. 
It clearly revels that switching is independent of the phase but depends only on polarization, being consistent with the picture of angular momentum transfer.

\clearpage
\section{Anisotropic magnetoresistance of Ni film.}
\begin{figure}[h]
  \centering
  \includegraphics[width = 0.8\linewidth]{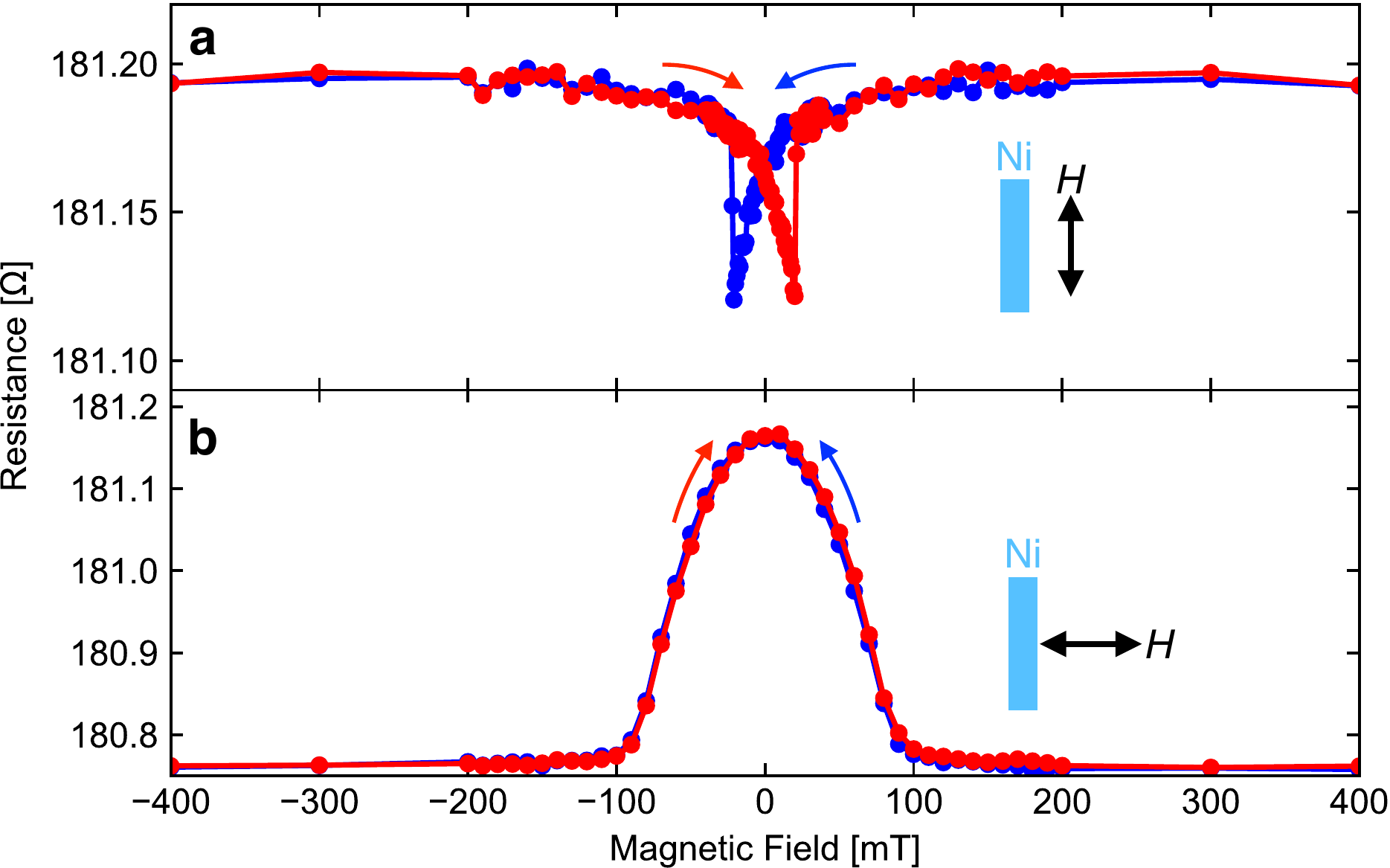}
  \caption{\textbf{Anisotropic magnetoresistance of Ni film.}
  \textbf{a}, The resistances in magnetic fields parallel to the Ni film. The blue and red line shows those in decreasing and increasing  magnetic fields. 
  \textbf{b}, The resistances in magnetic fields perpendicular to the Ni film. The blue and red line shows those in decreasing and increasing magnetic fields.}
  \label{fig:AMR}
\end{figure}

\clearpage
\section{Magnetization after the poling with the various magnetic field angles and SAW excitation magnitudes as probed by magnetoresistance}
\begin{figure}[h]
  \centering
  \includegraphics[width = \linewidth]{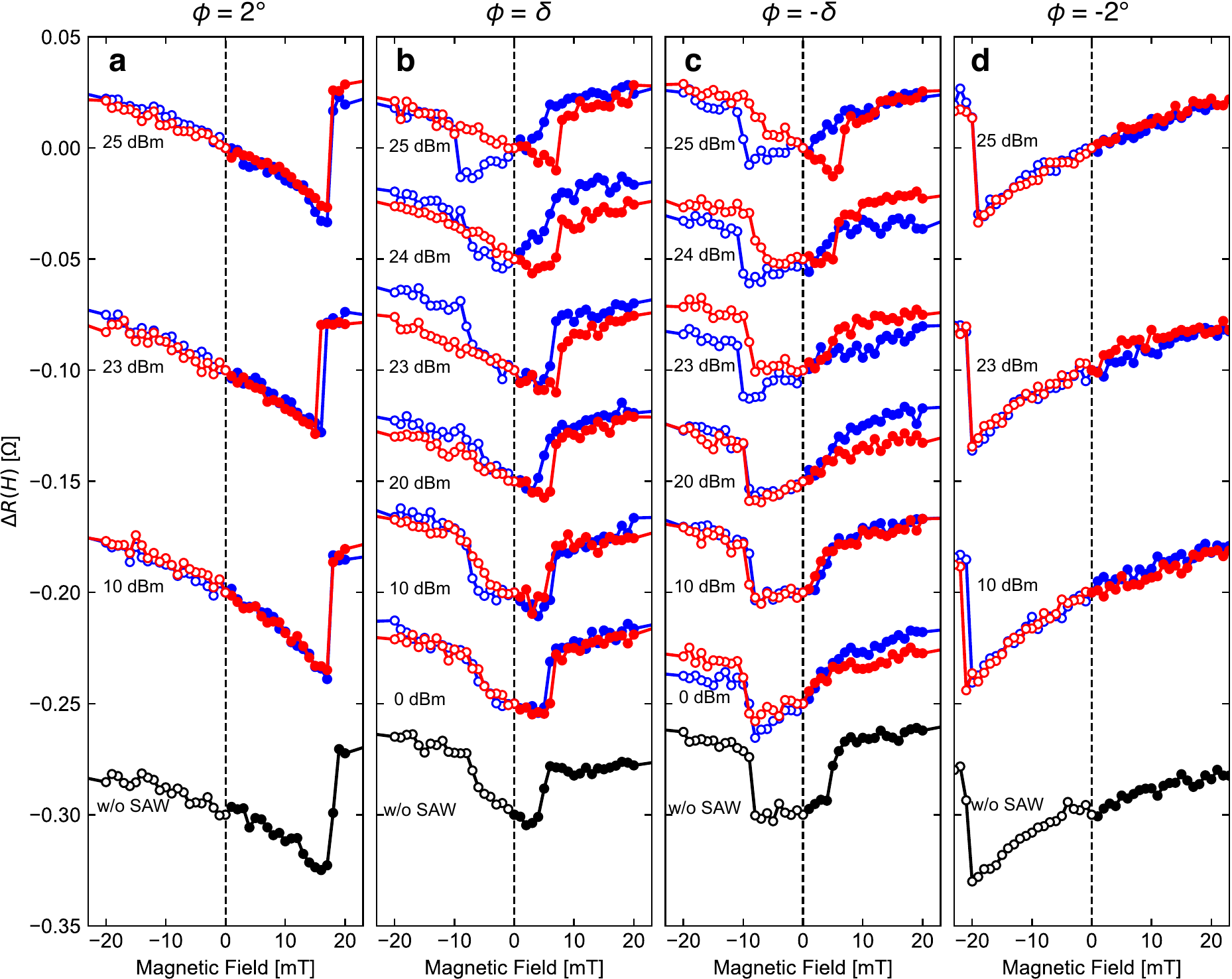}
  \caption{\textbf{Magnetoresistance after poling with various magnetic field angles and SAW excitation intensities.} 
  \textbf{a,b,c,d} Magnetoresistance $\Delta R(H) = R(H) - R(0)$ after the poling with various conditions. In the poling procedure, the magnetic field as large as 400 mT is initially applied, and SAW current is injected along $x$ or $-x$ direction. Then the magnitude of magnetic field is decreased to 0 T.
  The angles between $x$-axis and poling magnetic field are $+2^\circ$, $+\delta$, $-\delta$, $-2^\circ$ in a, b, c, and d, respectively. 
  The blue and red lines represent the magnetoresistance after the poling with SAW current along $+x$ and $-x$, respectively. The black dashed line emphasizes the initial field of 0 T. The data presented in the main text is that of $\phi=\delta$ and 25 dBm.
  }
  \label{fig:power dep}
\end{figure}

Figure \ref{fig:power dep} shows the magnetoresistance $\Delta R(H) = R(H) - R(0)$ after the poling with various magnetic field angles $\phi$ and SAW excitation magnitudes. 
At $\phi=\pm$ 2$^\circ$, the magnetization direction after the poling is determined only by the tilting direction of poling magnetic field, being insensitive to the direction and magnitude of SAW currents. The slope of magnetoresistance for $\phi = +2^\circ$  ($\phi=-2^\circ$) is negative (positive) and the discontinuous increase is observed at a positive (negative) magnetic field, indicating that the magnetization after the poling is along $+z$ ($-z$) direction. At $\phi=+\delta$ and $-\delta$ the magnetoresistance slightly decrease and increase around $H$ = 0 T, respectively, and does not show any SAW current direction dependence in the cases of small SAW excitation. When the SAW excitation increases above 23 dBm, the magnetoresistance becomes dependent on SAW current direction. At 25 dBm, the slope of magnetoresistance seems governed by the sign of SAW direction.

\clearpage
\section{Probing magnetization direction by nonreciprocity}

\begin{figure}[h]
  \centering
  \includegraphics[width = 0.4\linewidth]{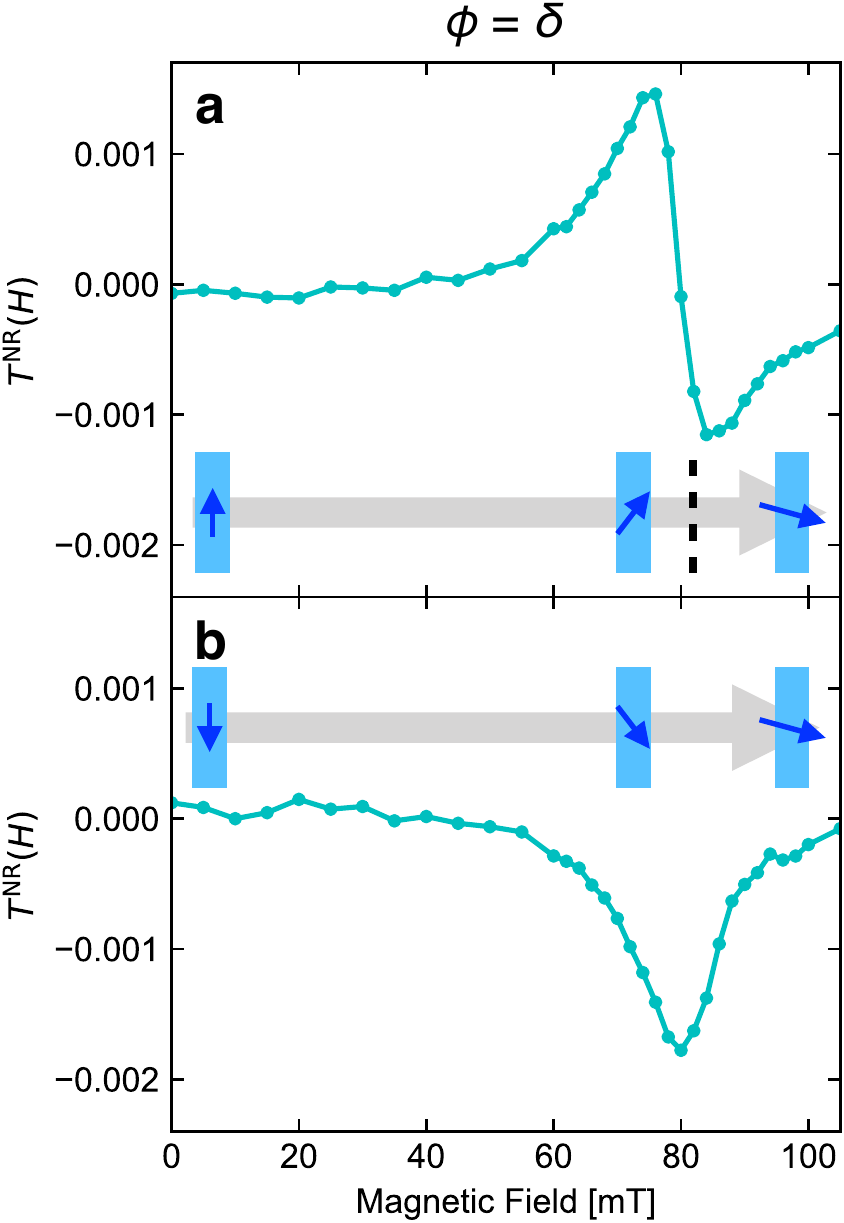}
  \caption{\textbf{Nonreciprocity after the application of the magnetic field parallel to the Ni film.}
  \textbf{a,b}, The magnetic field dependence of the nonreciprocity at $\phi = \delta$ after the application of magnetic fields as large as 400 mT parallel to $-z$ (a) and $+z$ axes (b) without any SAW excitation. Insets illustrate the inferred magnetization directions. The black dashed line emphasize the magnetization $\theta$ flop.
  }
  \label{fig:probe by NR}
\end{figure}

After a strong magnetic field is applied parallel (antiparallel) to the $z$-axis and decreased to zero, the magnetization should point at $+z$ ($-z$) direction. By using these states, we confirm that the measurement of SAW nonreciprocity certainly probe the magnetization direction after the poling.
Figures \ref{fig:probe by NR}a and b show variation of the nonreciprocity $T^{\rm NR}(H)$ in increasing the magnetic field at $\phi = \delta$ from 0 mT after the application and removal of magnetic fields as large as 400 mT antiparallel and parallel to $z$-axes, respectively. In measuring $T^{\rm NR}(H)$, the magnetic field is increased from 0 mT, and the microwave power is 10 dBm.
$T^{\rm NR}(H)$ increases and decreases for the initial magnetization antiparallel and parallel to $+z$-axis, respectively. 
The sign change due to the magnetization $\theta$ flop was discerned only in Fig. \ref{fig:probe by NR}a. These results confirm that the magnetic field variation of nonreciprocity certainly probe the direction of the magnetization at zero magnetic field.

\begin{figure}[t]
  \centering
  \includegraphics[width = 0.4\linewidth]{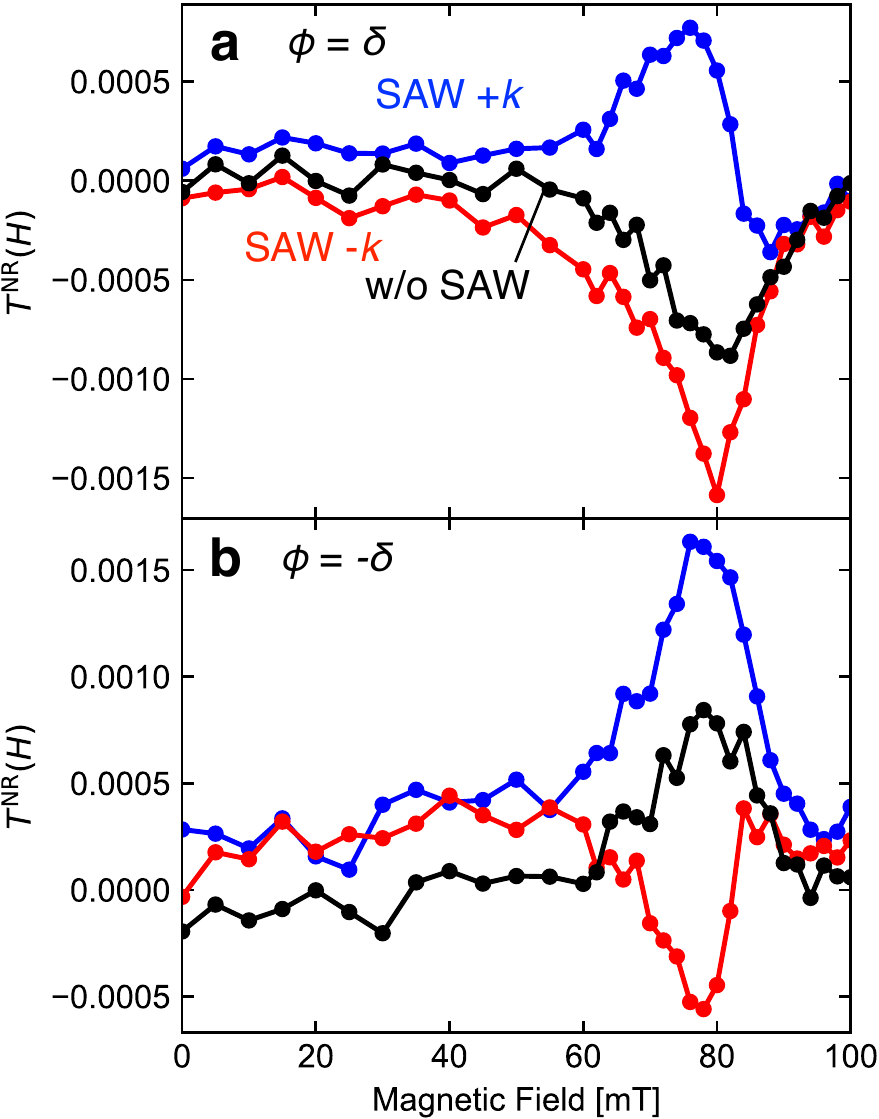}
  \caption{\textbf{Nonreciprocity after the poling with SAW current.}
  \textbf{a,b}, The magnetic field dependence of the nonreciprocity $T^{\rm NR}(H)$ after the poling with the magnetic fields at $\phi = \delta$ (\textbf{a}) and $-\delta$ (\textbf{b}). The blue and red lines represent $T^{\rm NR}(H)$ after the poling with SAW currents parallel and antiparallel to the $x$-axis, respectively. The black lines show those after the poling without the SAW input.
  }
  \label{fig:probe by NR 2}
\end{figure}

Figures \ref{fig:probe by NR 2}a and b show the magnetic field dependence of the nonreciprocity after the poling with or without SAW. In these cases, the poling and measurement magnetic fields have the same direction: $\phi=+\delta$ in Fig. \ref{fig:probe by NR 2}a and $\phi=-\delta$ in Fig. \ref{fig:probe by NR 2}b. 
The magnetization direction after the poling without the SAW (black line) seems determined by the tilted direction of the magnetic field in the poling process. 
On the other hand, in the presence of SAW current along $+x$ (blue line) and $-x$ (red line) directions, the magnetization direction seems governed by the SAW propagation direction in the poling procedure.

\bibliographystyle{naturemag}
\bibliography{bibliography}